\documentclass[twocolumn]{aastex62}

\graphicspath{{./}{figures/}}

\received{}
\revised{}
\accepted{ApJL}

\shorttitle{Overmassive black holes in dwarf galaxies}
\shortauthors{Mezcua, M. et al.}

\begin{document}

\title{Overmassive black holes in dwarf galaxies out to z$\sim$0.9 in the VIPERS survey}

\correspondingauthor{Mar Mezcua}
\email{marmezcua.astro@gmail.com}

\author[0000-0003-4440-259X]{Mar Mezcua}
\affiliation{Institute of Space Sciences (ICE, CSIC), Campus UAB, Carrer de Magrans, 08193 Barcelona, Spain}
\affiliation{Institut d'Estudis Espacials de Catalunya (IEEC), Carrer Gran Capit\`{a}, 08034 Barcelona, Spain}

\author{Malgorzata Siudek}
\affiliation{Institute of Space Sciences (ICE, CSIC), Campus UAB, Carrer de Magrans, 08193 Barcelona, Spain}
\affiliation{Institut de F\'isica d'Altes Energies, The Barcelona Institute of Science and Technology, 08193 Bellaterra, Spain}

\author{Hyewon Suh}
\affiliation{Gemini Observatory/NSF's NOIRLab, 670 N. A'ohoku Place, Hilo, HI 96720, USA}

\author{Rosa Valiante}
\affiliation{INAF-Osservatorio Astronomico di Roma, via di Frascati 33,00078 Monte Porzio Catone, Italy}
\affiliation{INFN, Sezione di Roma I, P.le Aldo Moro 2, 00185 Roma, Italy}

\author{Daniele Spinoso}
\affiliation{Department of Astronomy, 6th floor, MongManWai Building, Tsinghua University, Beijing 100084, China}

\author{Silvia Bonoli}
\affiliation{Donostia International Physics Center (DIPC), Manuel Lardizabal Ibilbidea, 4, Donostia-San Sebasti\'an, Spain}
\affiliation{IKERBASQUE, Basque Foundation for Science, E-48013, Bilbao, Spain}



\begin{abstract}
Supermassive black holes (SMBHs) are thought to originate from early Universe seed black holes of mass $M_\mathrm{BH} \sim 10^2$-10$^5$ M$_{\odot}$ and grown through cosmic time. Such seeds could be powering the active galactic nuclei (AGN) found in today's dwarf galaxies. However, probing a connection between the early seeds and local SMBHs has not yet been observationally possible. Massive black holes hosted in dwarf galaxies at intermediate redshifts, on the other hand, may represent the evolved counterparts of the seeds formed at very early times. 
We present a sample of seven broad-line AGN in dwarf galaxies with a spectroscopic redshift ranging from z=0.35 to z=0.93. The sources are drawn from the VIPERS survey as having a stellar mass ($M_\mathrm{*}$) LMC-like derived from spectral energy distribution fitting and they are all star-forming galaxies. Six of these sources are also X-ray AGN. The AGN are powered by SMBHs of $>10^7$ M$_{\odot}$, more massive than expected from the $M_\mathrm{BH}$-$M_\mathrm{*}$ scaling relation of AGN. Based on semi-analytical simulations, we find that these objects are likely overmassive with respect to their hosts since early times (z$>$4), independently of whether they formed as heavy ($\rm \sim 10^5$ M$_\odot$) or light ($\rm \sim 10^2$ M$_\odot$) seed black holes. In our simulations, these objects tend to grow faster than their host galaxies, contradicting models of synchronized growth. The host galaxies are found to possibly evolve into massive systems by z$\sim$0, indicating that local SMBHs in massive galaxies could originate in dwarf galaxies hosting seed black holes at higher z.
\end{abstract}



\section{Introduction} 
\label{intro}
The seed black holes from which SMBHs grow are expected to form at z$>$10 via direct collapse of primordial gas, mergers in dense stellar clusters, or the death of the first generation of Population III stars, among other possibilities (see \citealt{2017IJMPD..2630021M}; \citealt{2020ARA&A..58..257G} for reviews). Those seed black holes that did not grow could be today powering the AGN with M$_\mathrm{BH} \lesssim 10^{6}$ M$_{\odot}$ found in local dwarf galaxies (e.g., \citealt{2013ApJ...775..116R}; \citealt{2018ApJ...863....1C}). However, proving any connection between the early Universe seed black holes and local SMBHs has been so far prevented by the scarce number of AGN dwarf galaxies spectroscopically identified at high redshifts. 

Using the \textit{Chandra} COSMOS Legacy survey (\citealt{2016ApJ...819...62C}), \cite{2018MNRAS.478.2576M} found a sample of 40 AGN dwarf galaxies (with stellar masses log $M_\mathrm{*} \leq$ 9.5 M$_{\odot}$) out to photometric z$\sim$2.4. Thirteen of these sources have confirmed spectroscopic redshifts, one of them at z = 0.505 (lid$\_$391) and with an X-ray luminosity of log L$_\mathrm{0.5-10 keV}$ = 43.2 erg s$^{-1}$ (\citealt{2018MNRAS.478.2576M}). An AGN dwarf galaxy at a similar spectroscopic redshift (z$\sim$0.56) and X-ray luminosity (L$_\mathrm{0.5-10 keV} \sim 10^{43}$ erg s$^{-1}$) was also found by \cite{2019ApJ...885L...3H} (HSC-XD 52). Based on short-timescale variability in Dark Energy Survey (DES) lightcurves, \cite{2022MNRAS.516.2736B} have recently identified a dwarf galaxy at a spectroscopic redshift z = 0.8194 and classified as a 'QSO' by the Sloan Digital Sky Survey (SDSS; J022305.3-042800.9). Based on the presence of broad H$\beta$ and MgII emission lines, the authors report a black hole mass of log M$_\mathrm{BH}$ = 6.4-6.6 M$_{\odot}$, slightly larger than those typically found at z $\lesssim$ 0.15 (e.g., \citealt{2013ApJ...775..116R}). 
The highest redshift sample of AGN dwarf galaxies comes from the VLA-COSMOS 3 GHz Large Project (\citealt{2017A&A...602A...1S}), in which \cite{2019MNRAS.488..685M} found 35 radio AGN dwarf galaxies out to (photometric) z$\sim$3.4. Four of such sources have confirmed spectroscopic redshifts, one at z = 1.18 and another at z = 1.82 (\citealt{2019MNRAS.488..685M}), which constitute the redshift record-holders for an AGN in a dwarf galaxy. 

\cite{2023MNRAS.518..724S} have recently identified a sample of 4,315 AGN dwarf galaxies at 0.5 $<$ z $<$ 0.9 using the VIMOS Public Extragalactic Redshift Survey (VIPERS), carried out with the VIMOS spectrograph on the $\sim$8m Very Large Telescope (\citealt{2018A&A...609A..84S}). Using correlations based on the luminosity of narrow [OIII] and H$\beta$ emission lines (i.e. \citealt{2019MNRAS.487.3404B}), the authors report a median black hole mass for their sample of log M$_\mathrm{BH}$ = 8.2 M$_{\odot}$. This suggests that the sources are overmassive with respect to black hole-galaxy scaling relations, even if these bend up at the low-mass end (e.g., \citealt{2017IJMPD..2630021M}; \citealt{2018ApJ...855L..20M}; \citealt{2018ApJ...864L...6P}; \citealt{2020ARA&A..58..257G}).

In this Letter we make use of the VIPERS survey to identify AGN in dwarf galaxies at z $\gtrsim$ 0.4 based on the detection of broad emission lines. This yields the detection of seven broad-line AGN (BLAGN) dwarf galaxies, for which we derive black holes masses of log M$_\mathrm{BH}$ = 7.6-8.7 M$_{\odot}$ indicative of overmassive black holes. We then perform semi-analytical simulations to investigate the origin of these sources. The sample and data analysis are described in Sect.~\ref{sample}. The results obtained are discussed in Sect.~\ref{results}. Final conclusions are provided in Sect.~\ref{conclusions}. We adopt a $\Lambda$-CDM cosmology with $H_{0}=70$ km s$^{-1}$ Mpc$^{-1}$, $\Omega_{\Lambda}=0.73$ and $\Omega_{m}=0.27$. 

\section{Sample and analysis} 
\label{sample}
Our parent sample is composed of 1,161 galaxies classified as BLAGN during the VIPERS validation process of more than $\sim$90,000 VIPERS spectra and covering a redshift range from z=0.1 to z=4.56. Each spectroscopic redshift has an assigned flag corresponding to its quality (\citealt{2014A&A...562A..23G}; \citealt{2018A&A...609A..84S}). More than 80\% (935 of 1,161) of the identified BLAGN have the highest confidence ($>$95\%) of redshift measurements, supported in majority (626 sources) by at least one broad emission line. For 394 galaxies the lines were found not to be significantly broad, indicating that they might not be an AGN. Nevertheless, we do not exclude these from the parent sample, as the secure final sample of BLAGN in dwarf galaxies is selected with further cuts (based on their stellar masses and emission line measurements). 

We fit the multiwavelength spectral energy distribution (SED) of the 1,161 BLAGN galaxies using a combination of galaxy and AGN templates in order to derive the integrated stellar mass and star formation rate (SFR) of each galaxy (\citealt{2019ApJ...872..168S}). Details on the SED fitting are provided in the Appendix~\ref{SED}. Dwarf galaxies are then selected as having an SED with at least nine data points and a best-fitted stellar mass of log $M_\mathrm{*} \leq$ 9.5 M$_{\odot}$ similar to that of the Large Magellanic Cloud (\citealt{2006lgal.symp...47V}), as typically considered in searches for AGN in dwarf galaxies (e.g., \citealt{2013ApJ...775..116R,2020ApJ...888...36R}; \citealt{2016ApJ...817...20M,2018MNRAS.478.2576M,2019MNRAS.488..685M}; \citealt{2020ApJ...898L..30M}). To account for the uncertainties inherent to the SED-fitted stellar mass, we further consider only those galaxies with an uncertainty of $\sim$0.5 dex or less in the stellar mass probability distribution function (PDF) and a narrow distribution not extending beyond a stellar mass upper limit of 10$^{10}$ M$_{\odot}$. This ensures that all the sources in our sample can be reliably classified as low-mass galaxies. Applying these cuts yields a final sample of seven BLAGN dwarf galaxies. The SED best fits and stellar mass PDF of one of them are shown in the Appendix~\ref{SED}. To obtain independent measures of the stellar masses we also use the SED fitting code CIGALE (\citealt{2019A&A...622A.103B}) and the mass-to-light ratio derived with the code pPXF (\citealt{2017MNRAS.466..798C}). In both cases the stellar masses obtained are consistent with low-mass galaxies (see Appendix~\ref{stellarmass}).

To derive the black hole mass of the BLAGN dwarf galaxies we fit their optical spectra using the multi-component spectral fitting code PyQSOFit (\citealt{2018ascl.soft09008G}). The code subtracts the continuum emission and then models the H$\beta$ and MgII emission lines, among others, using a combination of narrow and broad Gaussian components. The uncertainties of the fitting results are derived using a MonteCarlo estimation. Single-epoch virial black hole masses are then derived from the H$\beta$ and MgII (depending on the redshift) broad-line width and adjacent continuum/line luminosity assuming that the gas is virialized. A detailed description of the emission line fitting and black hole mass estimation is provided in Appendix~\ref{broadlines} and Appendix~\ref{bhmass}.

\section{Results and discussion} 
\label{results}
The spectroscopic redshifts for the seven BLAGN dwarf galaxies range from z=0.35 to z=0.93, which constitute the highest confirmed redshifts for a sample of BLAGN in dwarf galaxies. With SFRs ranging from log SFR = 0.4 to 1.5 M$_{\odot}$ yr$^{-1}$ and a mean value of log SFR = 1.1 M$_{\odot}$ yr$^{-1}$, the galaxies sit above the main-sequence of star-forming galaxies (e.g., \citealt{2014ApJ...795..104W}), which proves their star-forming nature.
Six of the sources have an XMM-Newton counterpart at 2-10 keV in the XMM-XXL catalogue (\citealt{2018A&A...620A..12C}), from which we derive a k-corrected luminosity L$_\mathrm{2-10 keV}$ = 10$^{43}$ - 10$^{44}$ erg s$^{-1}$ that confirms their AGN nature. For one of them, 106174409, the spectral fitting performed in \cite{2016A&A...592A...5F} yields an intrinsic column density of N$_{H}$ = 3.1 $\times$ 10$^{19}$ cm$^{-2}$, consistent with the low level of absorption of a BLAGN. For 106174409, as well as for 106158702 and 110167802, we are also able to derive the Balmer decrement in terms of H$\alpha$/H$\beta$ and H$\gamma$/H$\beta$. We find Balmer decrements that are consistent with the theoretical ones, indicating that intrinsic absorption is not significant and supporting the broad-line nature of these sources. Table 1 provides the most relevant information of each AGN dwarf galaxy in the sample. 

\begin{table*}
\begin{center}
\caption{Properties of the sample of seven BLAGN dwarf galaxies.}
\label{table1}
\begin{tabular}{ccccccccc}
\hline
\hline 
VIPERS &  $z$ &  log $M_\mathrm{*}$ & log $M_\mathrm{BH}$  & log $L_\mathrm{bol}$  & $\lambda_\mathrm{Edd}$ & Survey & Broad& log $L_\mathrm{X}$  \\
ID 	     & 		& (M$_{\odot}$)          & (M$_{\odot}$)                 & (erg s$^{-1}$)               &					&              & Line & (erg s$^{-1}$) \\
(1) & (2)   & (3)   & (4)   & (5)   & (6)   & (7)   &  (8) & (9) \\
\hline
106174409  & 0.354  & 9.2$\pm$0.3 & 8.4$\pm$0.3 & 44.9  & 0.02 & SDSS & H$\beta$ &  43.6$\pm$0.1  \\
120102612  & 0.578  & 8.6$\pm$0.4 & 8.1$\pm$0.3 & 44.8  & 0.04 & VIPERS & H$\beta$ &  43.4$\pm$0.2  \\
106158702  & 0.599  & 9.5$\pm$0.3 & 7.6$\pm$0.4 & 45.4  & 0.5 &VIPERS & H$\beta$ &  43.4$\pm$0.5  \\
114143514  & 0.628  & 9.4$\pm$0.3 & 8.1$\pm$0.3 & 45.1  & 0.08 & VIPERS & H$\beta$ &  43.4$\pm$0.2  \\
110167802  & 0.671  & 9.0$\pm$0.4 & 7.9$\pm$0.3 & 45.1  & 0.1 & VIPERS & H$\beta$ &  43.6$\pm$0.5  \\
103138295  & 0.900  & 9.3$\pm$0.6 & 7.6$\pm$0.5 & 45.3  & 0.4 & SDSS & MgII &  --  \\
117047934  & 0.928  & 9.2$\pm$0.7 & 8.7$\pm$0.3 & 45.3  & 0.04 & SDSS & MgII &  44.2$\pm$0.3  \\
\hline
\hline
\end{tabular}
\end{center}
\smallskip\small {\bf Column designation:}~(1) VIPERS ID; (2) redshift; (3) stellar mass derived from SED fitting; (4) black hole mass derived from single-epoch virial calibrations; (5) bolometric luminosity derived from the monochromatic continuum luminosity at 3000 or 5100 \AA; (6) Eddington ratio; (7) survey for spectroscopy; (8) fitted broad emission line (9) 2-10 keV X-ray luminosity, if available. The uncertainties in the stellar mass include a 0.2 dex to account for differences in the stellar population models. The uncertainties in black hole mass are the quadratic sum of the measurement uncertainties ($\sim$0.1 dex) and the systematic uncertainties carried by single-epoch virial calibrations ($\sim$0.3 dex).
\end{table*}

\subsection{Black hole mass and Eddington rate}
We find a range of black holes masses for the seven BLAGN dwarf galaxies of log $M_\mathrm{BH}$ = 7.6-8.7 M$_{\odot}$ with an average uncertainty of 0.4 dex, fully consistent with SMBHs. The bolometric luminosities of the BLAGN dwarf galaxies, derived from the monochromatic continuum luminosity (\citealt{2006ApJS..166..470R}; see Appendix~\ref{bhmass}), range from log L$_\mathrm{bol}$ = 44.8 to 45.4 erg s$^{-1}$. These high luminosities are consistent with those of type 1 quasars (e.g., \citealt{2006ApJS..166..470R}) and are several orders of magnitude higher than those of local AGN dwarf galaxies (log L$_\mathrm{bol} \sim$ 40-42 erg s$^{-1}$; \citealt{2013ApJ...775..116R}; \citealt{2020ApJ...898L..30M}). Two of the sources are indeed catalogued as quasars in the SDSS (\citealt{2019ApJS..244...36C}). From the black hole mass and bolometric luminosity we derive a range of Eddington ratios $\lambda_\mathrm{Edd}$ = 0.02 - 0.5 with a mean value $\lambda_\mathrm{Edd}$ = 0.2, which indicates that the sources are mostly accreting at sub-Eddington rates. 

\subsection{The $M_\mathrm{BH}$-$M_\mathrm{*}$ scaling relation}
\begin{figure}
\centering
\includegraphics[width=0.47\textwidth]{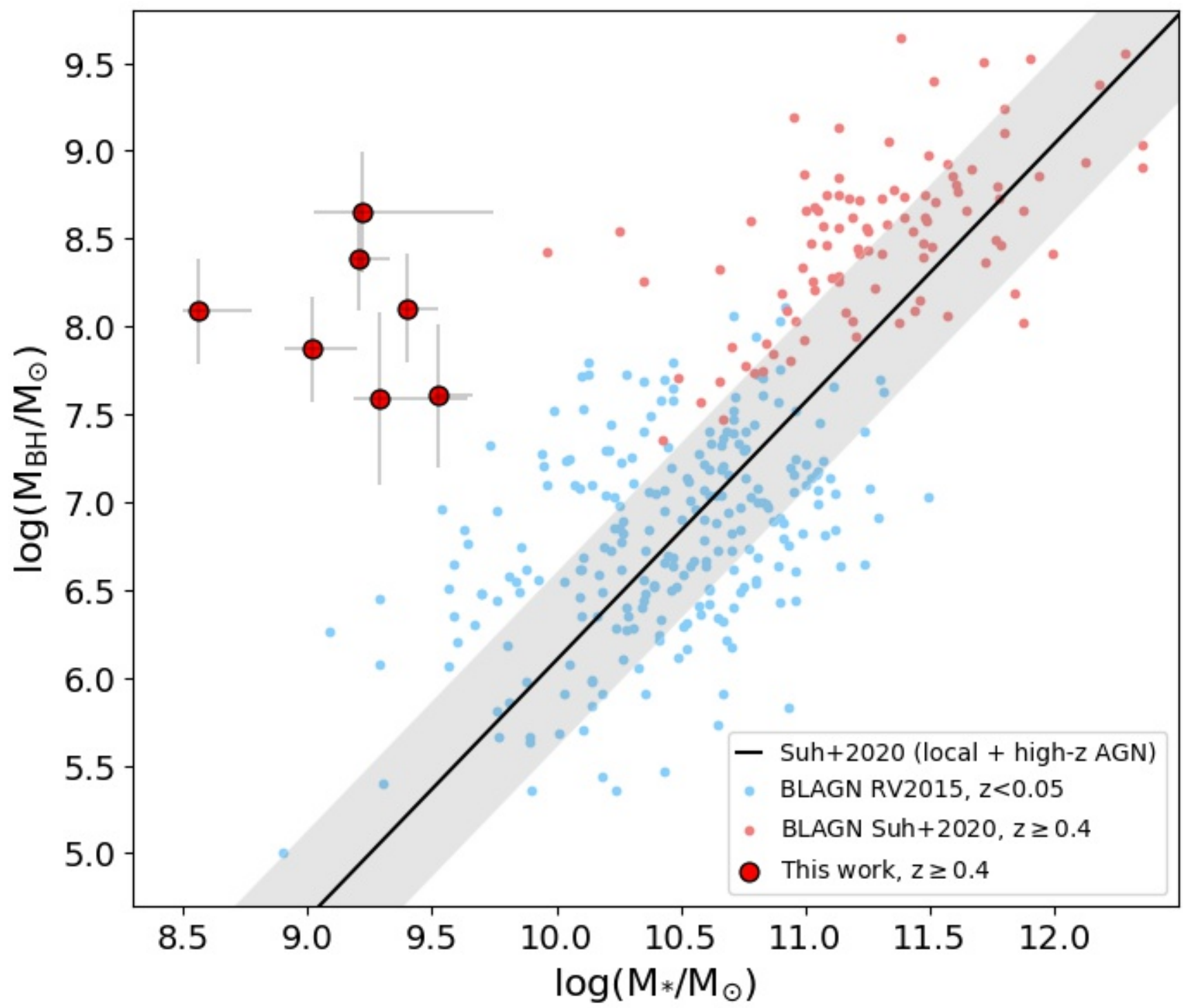}
\caption{$M_\mathrm{BH}$ versus $M_\mathrm{*}$ for the VIPERS sample of BLAGN dwarf galaxies. We show for comparison the low-z AGN of \cite{2015ApJ...813...82R} (RV2015) and the high-z AGN of \cite{2020ApJ...889...32S} (Suh+2020), whose masses have been computed using the same procedure and parameters as in our sample. The solid line shows the local + high-z $M_\mathrm{BH}$-$M_\mathrm{*}$ correlation found for the combination of the RV2015 and Suh+2020 samples with a 1$\sigma$ scatter of 0.5 dex.}
\label{MbhMstellar}
\end{figure}

To further investigate the sample of BLAGN dwarf galaxies we plot them on a $M_\mathrm{BH}$-$M_\mathrm{*}$ diagram (Fig.~\ref{MbhMstellar}) and compare their location to that of local (z$<$0.05) BLAGN in both dwarf and massive galaxies (\citealt{2015ApJ...813...82R}; from now on RV2015 sample) and to that of high-z (z $\geq$ 0.4) BLAGN in massive galaxies (\citealt{2020ApJ...889...32S}; from now on Suh+2020 sample). The black hole masses of all these samples are derived using a similar virial factor as our value and the stellar masses corrected for any possible changes in the mass-to-light ratios (see Appendix~\ref{bhmass}). We also plot in Fig.~\ref{MbhMstellar} the local + high-z $M_\mathrm{BH}$-$M_\mathrm{*}$ correlation found for the combination of the RV2015 and Suh+2020 samples (\citealt{2020ApJ...889...32S}). We find that all the BLAGN in our sample of dwarf galaxies are overmassive with respect to the local + high-z $M_\mathrm{BH}$-$M_\mathrm{*}$ correlation of \cite{2020ApJ...889...32S}. The median of the black hole mass offset computed using a MonteCarlo approach (see Appendix~\ref{bhmass}) is of $\Delta M_\mathrm{BH}$= 3.2 $\pm$ 1.3 with a significance of 100\% (3$\sigma$ level).

\begin{figure}
\centering
\includegraphics[width=0.47\textwidth]{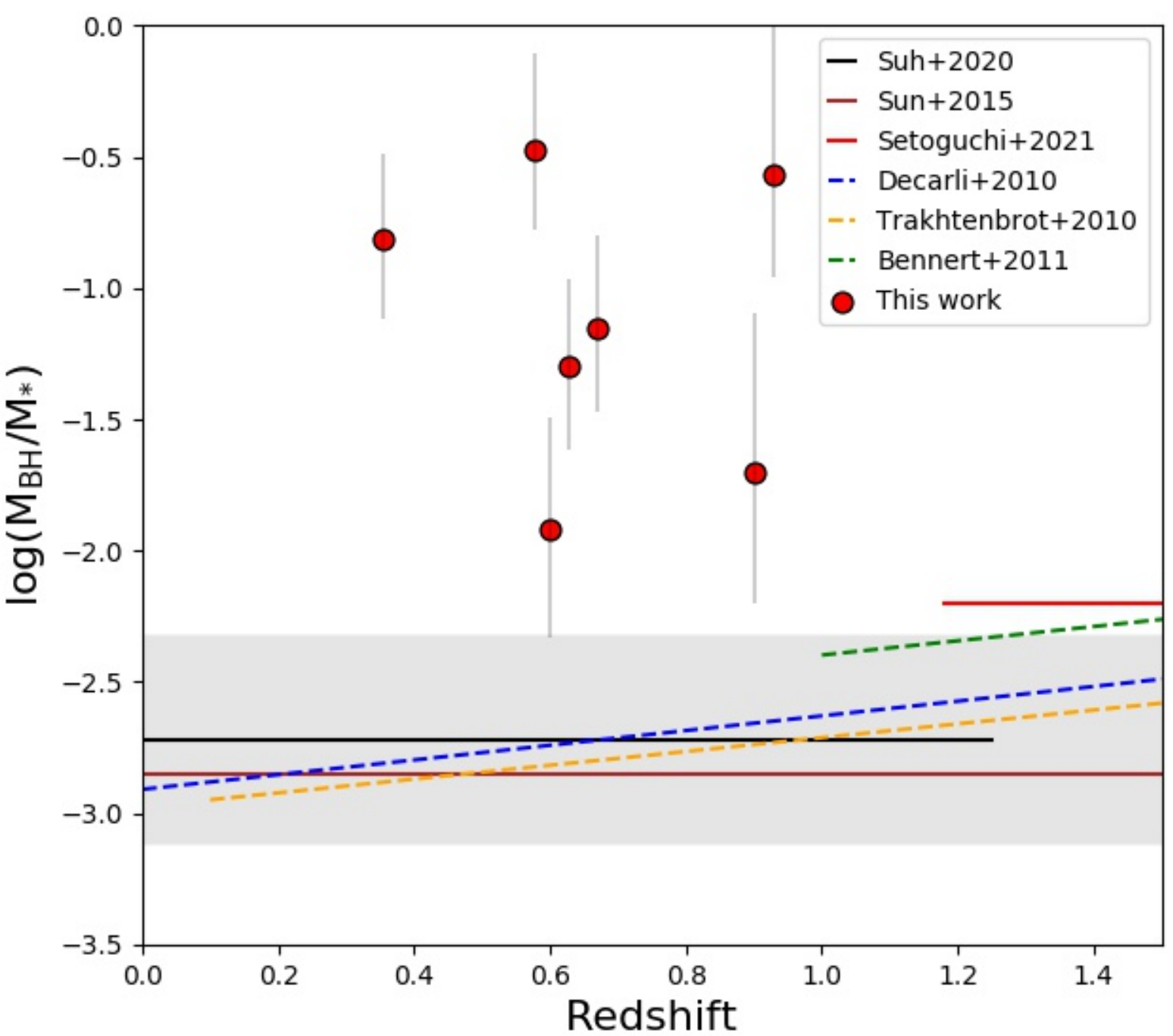}
\caption{Ratio of $M_\mathrm{BH}$/$M_\mathrm{*}$ versus redshift for the sample of BLAGN dwarf galaxies. We show for comparison the non-evolution found by \cite{2020ApJ...889...32S} for z$\sim$0-2.5 (black line, 1$\sigma$ scatter$\sim$0.5 dex), \cite{2015ApJ...802...14S} for z$\sim$0-2 (brown line), and \cite{2021ApJ...909..188S} for z$\sim$1.2-1.7 (red line), and the z-evolution found by \cite{2010MNRAS.402.2453D} for z$\sim$0-3 (dashed blue line), \cite{2010MNRAS.406L..35T} for z$\sim$0.1-2  (dashed yellow line), and \cite{2011ApJ...742..107B} for z$\sim$1-2 (dashed green line, including the data of \citealt{2010ApJ...708..137M}). We note that the Suh+2020 sample is the only one for which the black hole and stellar masses have been computed using the same procedure and parameters as in our sample.}
\label{MbhMstellarevolution}
\end{figure}

\cite{2020ApJ...889...32S} find no redshift evolution of the $M_\mathrm{BH}$-$M_\mathrm{*}$ correlation out to z$\sim$2.5, in agreement with most studies (e.g., \citealt{2009ApJ...706L.215J}; \citealt{2011ApJ...741L..11C}; \citealt{2012ApJ...753L..30M}; \citealt{2015ApJ...802...14S}; \citealt{2021ApJ...909..188S}); however, the opposite is found in some other works (e.g., \citealt{2010MNRAS.402.2453D}; \citealt{2010ApJ...708..137M}; \citealt{2010MNRAS.406L..35T}; \citealt{2011ApJ...742..107B}; \citealt{2020ApJ...888...37D}). In Fig.~\ref{MbhMstellarevolution} we show the evolution of the $M_\mathrm{BH}$/$M_\mathrm{*}$ ratio with redshift found in some of these studies. We see that all the black holes in our BLAGN sample are outliers, defined as having $M_\mathrm{BH} > 10^7$ M$_{\odot}$ and $M_\mathrm{BH}$/$M_\mathrm{*} >$ 0.01 (\citealt{2019MNRAS.485..396V}), independently of whether the $M_\mathrm{BH}$/$M_\mathrm{*}$ ratio evolves with redshift or not. The presence of $M_\mathrm{BH}$/$M_\mathrm{*}$ outliers has been found in massive galaxies in the local Universe (e.g., \citealt{2015ApJ...808...79F}; \citealt{2016ApJ...817....2W}) and at z$\sim$3 (e.g., \citealt{2015Sci...349..168T}) and can be explained if the black holes have grown more efficiently than their host galaxies, contrarily to models of synchronized growth. In the low-mass regime, both compact elliptical galaxies and ultracompact dwarves are found to often host central SMBHs way more massive than expected from their hosts (e.g., \citealt{2021MNRAS.506.4702F}; \citealt{2014Natur.513..398S}; \citealt{2018MNRAS.477.4856A}). While the most likely origin of such outliers is tidal stripping caused by their location in groups and clusters (e.g., \citealt{2018MNRAS.473.1819F,2021MNRAS.506.4702F}), none of our BLAGN dwarf galaxies resides in such a high-density environment (see Appendix~\ref{environment}). A few other overmassive black holes have been also found in dwarf galaxies in the local Universe (e.g., Was 49b, \citealt{2017ApJ...836..183S}), however no such results had been so far observationally reported for a sample of AGN dwarf galaxies at intermediate redshifts (z $\sim$ 0.4-0.9). Such a population of overmassive black holes in dwarf galaxies has been recently found in cosmological numerical simulations at z=0-2 (e.g., figures 3 and 4 in \citealt{2021MNRAS.503.1940H}) and beyond (e.g., \citealt{2021MNRAS.503.3568K}).

\subsection{Semi-analytical simulations}
To investigate on the possible origin of the BLAGN dwarf galaxies we perform cosmological semi-analytical simulations to trace the history of these sources. We make use of the updated version of the L-Galaxies semi-analytical model (\citealt{2023MNRAS.518.4672S}) and apply it to the merger trees of the Millenium II cosmological, N-body simulation (\citealt{2009MNRAS.398.1150B}). This model allows to follow the joint evolution of galaxies and their massive black holes, whose first seeds form through four different channels, ranging from light PopIII remnants with mass $\sim$100 M$_{\odot}$, up to the heavy direct-collapse black holes of $\sim10^5$ M$_{\odot}$ (\citealt{2023MNRAS.518.4672S}). In particular, due to the mass-resolution limits of the Millenium II simulation, we cannot model directly the formation and evolution of PopIII stars, therefore we inherit the evolved population of light-seeds from the GQd semi-analytical cosmological simulation (\citealt{2016MNRAS.457.3356V,2021MNRAS.500.4095V}).

More in detail, light-seeds are obtained in GQd as remnants of PopIII stars, whose formation and evolution is modeled according to physically motivated prescriptions (see for details \citealt{2016MNRAS.457.3356V}). Light-seeds in GQd grow in mass via gas-accretion and black hole-black hole mergers during their evolution (un-resolved by L-Galaxies on the Millennium-II merger trees). As detailed in \cite{2023MNRAS.518.4672S}, these evolved light-seeds descendants are then used as initial conditions (or "initial black hole seeds") in the L-Galaxies model, as soon as a new dark matter halo is resolved in the Millennium-II merger trees. In particular, after matching in both redshift and virial mass the newly-resolved dark matter halos of the Millennium-II to GQd structures, the evolved light-seeds (eventually) hosted by the latter are used as initial black holes in the L-Galaxies model. Finally, on top of this \textit{grafting} of light-seeds descendants, the eventual formation of intermediate-mass or heavy black hole seeds is also modeled self-consistently in L-Galaxies (see for further details \citealt{2023MNRAS.518.4672S}). During their evolution, all black holes formed in L-Galaxies (or inherited from GQd) grow in mass via gas accretion and/or black hole-black hole mergers following the model described in \cite{2020MNRAS.495.4681I}. This procedure extends and complements the outputs of the GQd model over the whole L-Galaxies/Millennium-II simulation, allowing to trace down to z=0 the cosmological evolution of massive black holes formed at z$>$6 through a physically-motivated black hole-seeding model. Furthermore, by using the L-Galaxies/Millennium II, we can follow this process over a dynamical range encompassing dwarf galaxies and Milky Way-type systems (i.e. 7 $\leq$ log $M_\mathrm{*}/M_{\odot} \leq$ 12).

Driven by the observational constraints derived in this work, we focus on the class of dwarf galaxies (i.e. log $M_\mathrm{*} \leq$ 9.5 M$_{\odot}$) that manage to host black holes with log $M_\mathrm{BH} \geq$ 7.6 M$_{\odot}$at 0.35 $\leq z \leq$ 1 and, among these galaxies, we look for systems which also show 0.4 $<$ log SFR (M$_{\odot}$ yr$^{-1}$) $<$ 1.5 and 0.02 $< \lambda_\mathrm{Edd} <$ 0.5 at the corresponding z. We find around 600 objects showing $M_\mathrm{*}$ and $M_\mathrm{BH}$ consistent with the observational limits when considering each L-Galaxies/Millenium II snapshot between z$\sim$0.35 and z$\sim$1. Among these candidates, a total of 30 objects (all at 0.35 $< z \leq$ 1) satisfy the observational constraints on either SFR or $\lambda_\mathrm{Edd}$ and one on both. All of them include black hole seed that were inherited from the GQd model as evolved light-seeds descendants of $M_\mathrm{BH} \sim 10^{3-4}$ M$_{\odot}$, hence effectively being already as massive as intermediate/heavy seeds in the L-Galaxies/Millennium II semi-analytical simulations. This leaves open the possibility that more than one black hole-formation channel can explain the formation of the overmassive black holes observed by this work.

\begin{figure}
\centering
\includegraphics[width=0.47\textwidth]{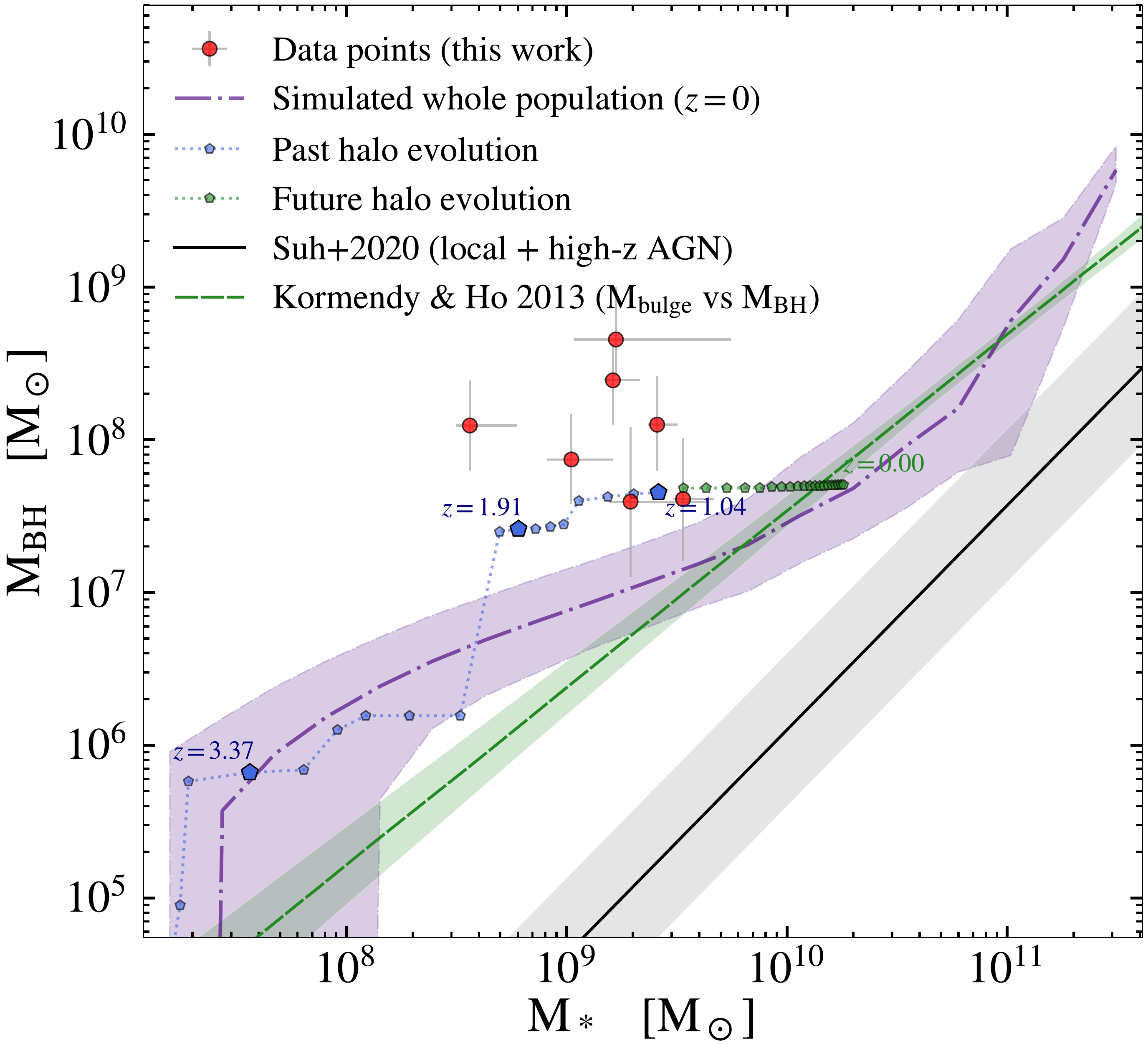}
\caption{Cosmological semi-analytical simulations of a plausible evolution history of one of the BLAGN dwarf galaxies. Its full history (blue pentagons and dotted line) shows that it is overmassive with respect to the $M_\mathrm{BH}$-$M_\mathrm{*}$ scaling relation of \citeauthor{2020ApJ...889...32S} (2020, grey solid line) since at least z $>$ 4. Extending the evolution of the simulations down to z=0 (green pentagons and dotted line), the candidate tends to reach a stellar mass $M_\mathrm{*} > 10^{10}$ M$_{\odot}$ and to move towards the local $M_\mathrm{BH}$-$M_\mathrm{bulge}$ relation of inactive massive early-type galaxies of \citeauthor{2013ARA&A..51..511K} (2013, green dashed line). The median $M_\mathrm{BH}$-$M_\mathrm{*}$ relation for all the z=0 galaxies simulated by L-Galaxies/Millennium II is shown as a purpled dashed line (with the purple area showing the 16th/84th percentiles of the distribution) . }
\label{sim}
\end{figure}

We show in Fig.~\ref{sim} the cosmological evolution of one of the 31 simulated objects that satisfy the observational constraints. By tracking the full history of these objects (light-blue pentagons and dotted line), we find that they tend to be overmassive with respect to the $M_\mathrm{BH}$-$M_\mathrm{*}$ scaling relation of \cite{2020ApJ...889...32S} over the cosmological evolution of their host galaxies (as traced within the L-Galaxies/Millennium II simulation). In particular, we find that they become overmassive mainly via a combination of black hole-black hole coalescence and gas-accretion episodes following the galaxy mergers experienced by their hosts. Indeed, the seed mass typically contributes by less than $\sim0.1\%$ to the total $M_\mathrm{BH}$ of these objects at 0.35$<z<$1. This shows that the main reason why these massive black holes are overmassive at $z<$1 has to be searched in their evolution rather than in their massive origin. Although the details of the evolution of each of the 31 simulated objects may differ, the average/typical picture we derive from our semi-analytical simulations is that efficient gas-accretion triggered by early (z$>$4) gas-rich mergers helps to push the evolved light-seeds inherited from GQd up to $M_\mathrm{BH} \gtrsim 10^6$ M$_\odot$ by $z\sim3$. Afterwards, the sequence of galaxy mergers experienced by their hosts induces these massive black holes coalescence with other massive black holes, hence "keeping" them overmassive along their evolution. Nevertheless, we note that this qualitative scenario should be considered with some caution, due to the high uncertainties associated to mass-growth model of black holes in the regime of dwarf galaxies sampled by L-Galaxies on the Millennium-II (as discussed in \citealt{2023MNRAS.518.4672S}).

On the other hand, the future evolution of these objects is variate. We find that 13\% of the 31 simulated objects remain hosted by dwarf $M_\mathrm{*} \lesssim 10^9$ M$_\odot$ systems and thus keep on being overmassive at $z$=0. Of the remaining sources, 35\% end up into ordinary, massive $M_\mathrm{*} > 10^{10}$ M$_{\odot}$ local galaxies, approaching both the median relation computed for all z=0 simulated galaxies (purple dashed-dotted line and shaded area in Fig.~\ref{sim}) as well as the local $M_\mathrm{BH}$-$M_\mathrm{bulge}$ relation of inactive early-type galaxies of \citeauthor{2013ARA&A..51..511K} (2013, green dashed line and shaded area). This is indeed the case of the representative example candidate shown in Fig.~\ref{sim} (green pentagons and dotted line). This suggests that approximately one third of today's SMBHs in massive galaxies could have their origin in higher-z dwarf galaxies that become massive later on.  
 
\section{Conclusions} 
\label{conclusions}
Black holes of $\sim10^{3}$-$10^{5}$ M$_{\odot}$ formed at $z >$ 10 are thought to be the seeds from which SMBHs grew. Such seed black holes should be found in local dwarf galaxies if they have not grown much through mergers and accretion (\citealt{2019NatAs...3....6M}), where they can be detected as AGN. In the last years several efforts have been made to detect such AGN dwarf galaxies out to z $\sim$ 1 (\citealt{2018MNRAS.478.2576M,2019MNRAS.488..685M}; \citealt{2019ApJ...885L...3H}; \citealt{2023MNRAS.518..724S}); however, the scarce number of high-z sources and for which a black hole mass estimate is available has prevented making a direct connection between them and the early Universe seed black holes. In this Letter we have reported the finding of a sample of seven AGN dwarf galaxies out to z $\sim$ 0.9 and with black hole masses $>10^{7}$ M$_{\odot}$ derived from the width of broad emission lines. The sources are overmassive with respect to the $M_\mathrm{BH}$-$M_\mathrm{*}$ scaling relation, which is expected from semi-analytical cosmological simulations including both PopIII remnants and direct collapse black hole seeds. This indicates that both pathways can reproduce the observational results, or that we cannot discern between seeding models even when going to redshifts higher than that of most of the AGN dwarf galaxy samples (z$<$0.1; e.g., \citealt{2013ApJ...775..116R}; \citealt{2018ApJ...863....1C}; \citealt{2020ApJ...898L..30M}). 

The descendants of the galaxies hosting overmassive black holes at 0.35 $< z <$1 in the simulation show a wide distribution of stellar masses, with 13\% of them having $M_\mathrm{*} \lesssim 10^9$ M$_\odot$ and hence continuing to host overmassive black holes at z=0 (as already found observationally in a few cases, e.g., \citealt{2017ApJ...836..183S}; \citealt{2022ApJ...937....7S}). On the other hand, 35\% of them (roughly one third) evolves into ordinary $M_\mathrm{*} > 10^{10}$ M$_{\odot}$ galaxies, hence becoming "normal" systems (i.e. massive galaxies hosting SMBHs). This indicates that dwarf galaxies hosting overmassive black holes at high-z could be the progenitors of (at least one third of) today's massive galaxies. The superb sensitivity of up-coming spectroscopic surveys such as DESI or the LSST will expand the number of BLAGN dwarf galaxies detected at high-z, allowing us to investigate further black hole evolution from the early seeds to local SMBHs. 

\acknowledgments
The authors thank the anonymous referee for helpful comments and suggestions. MM acknowledges support from the Ram\'on y Cajal fellowship (RYC2019-027670-I) and from the Spanish Ministry of Science and Innovation through the project PID2021-124243NB-C22. MS acknowledges support from the Polish National Agency for Academic Exchange (Bekker grant BPN/BEK/2021/1/00298/DEC/1), the European Union's Horizon 2020 research and innovation programme under the Maria Sklodowska-Curie (grant agreement No 754510) and by the Spanish Ministry of Science and Innovation through the Juan de la Cierva-formaci\'on program (FJC2018-038792-I). HS is supported by the international Gemini Observatory, a program of NSF's NOIRLab, which is managed by the Association of Universities for Research in Astronomy (AURA) under a cooperative agreement with the National Science Foundation, on behalf of the Gemini partnership of Argentina, Brazil, Canada, Chile, the Republic of Korea, and the United States of America. S.B. acknowledges support from the Spanish Ministry of Science and Innovation through the project PID2021-124243NB-C21. This work was also partially supported by the program Unidad de Excelencia Mar\'ia de Maeztu CEX2020-001058-M.



\bibliographystyle{aasjournal}
\bibliography{/Users/mmezcua/Documents/referencesALL}

\clearpage
\appendix

\section{Spectral energy distribution}
\label{SED}
\begin{figure}
\centering
\includegraphics[width=0.99\textwidth]{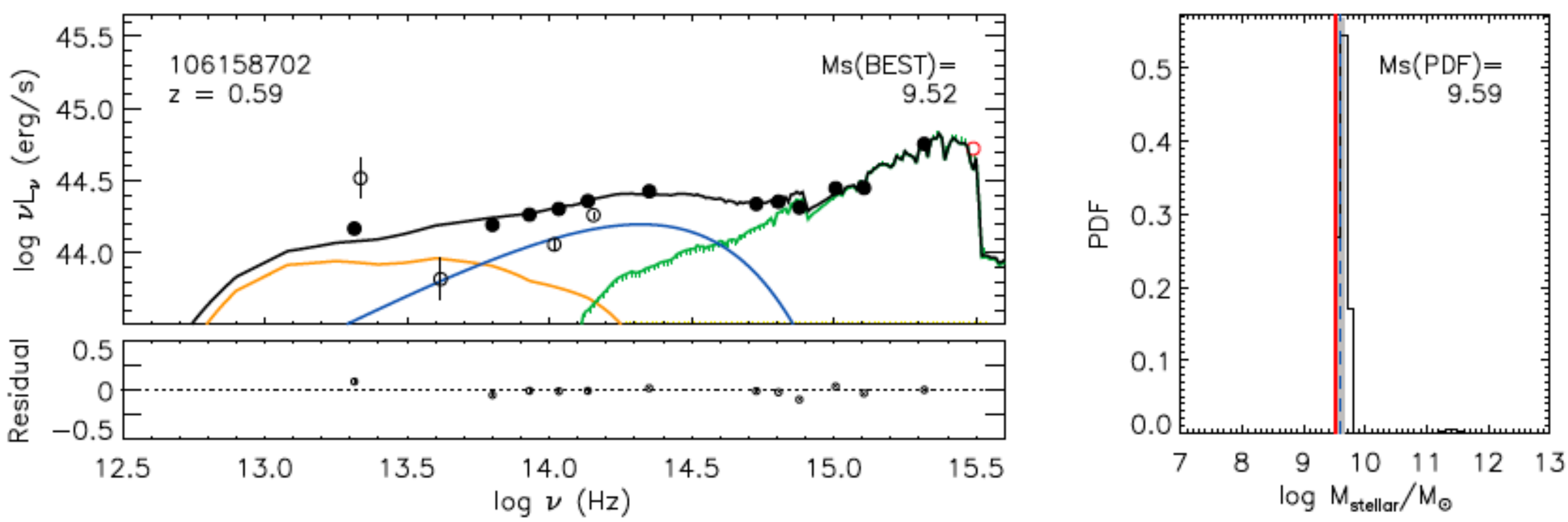}
\includegraphics[width=0.99\textwidth]{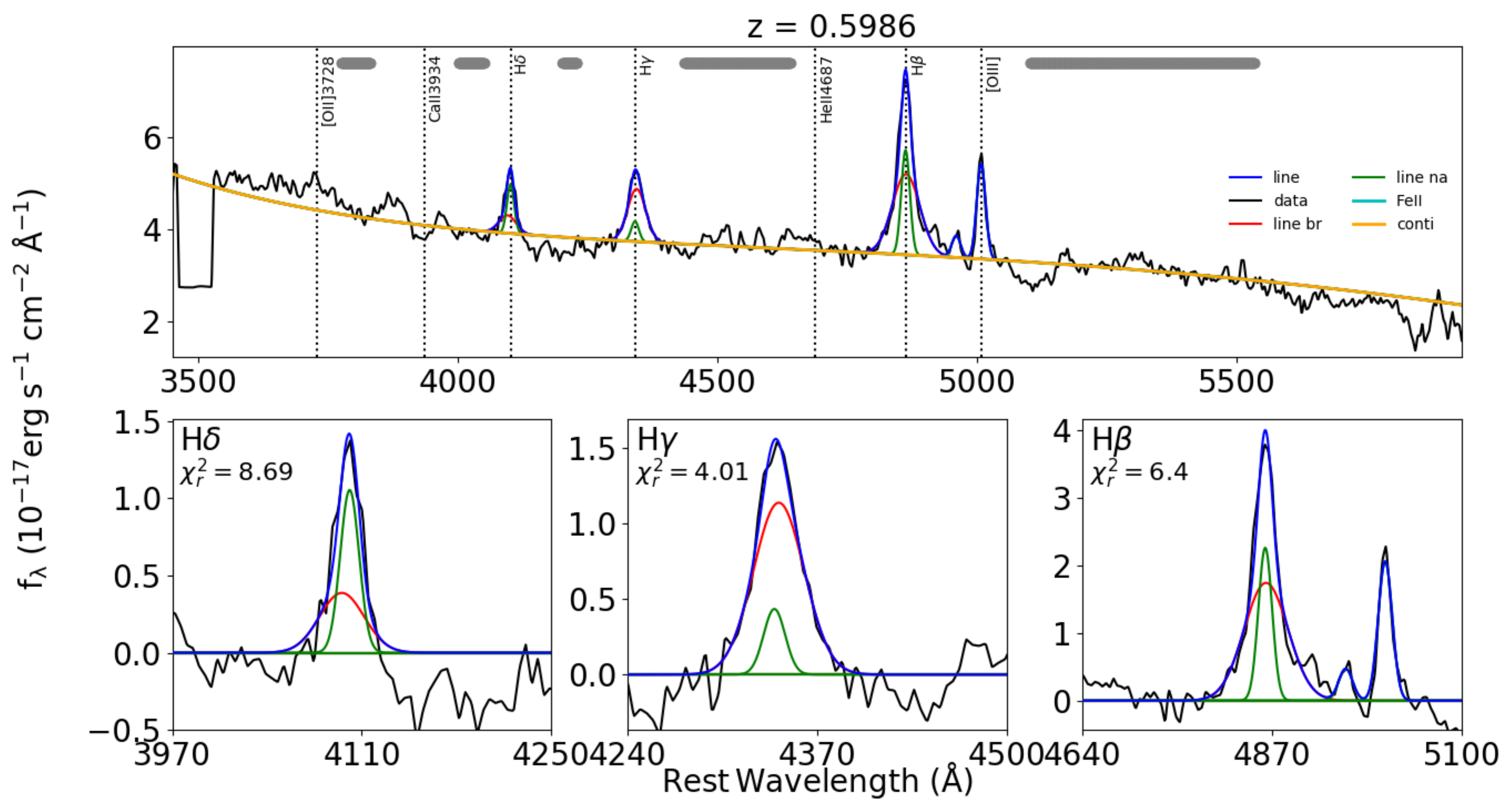}
\includegraphics[width=0.3\textwidth]{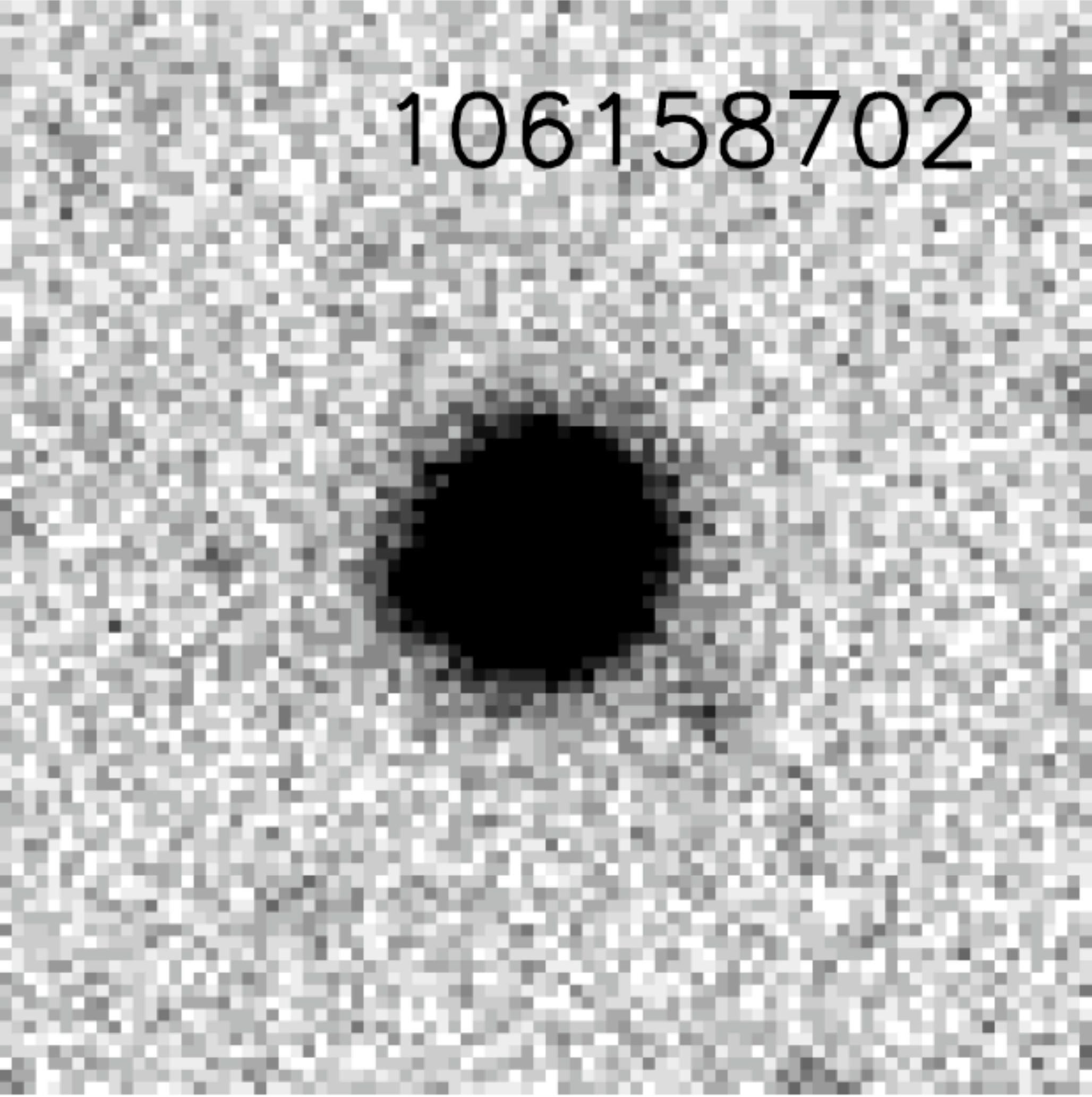}
\caption{Top left: Spectral energy distribution fitting of the rest-frame observed ultraviolet, optical, and infrared (when available) photometry (black points) with the best-fit model (black curve), including a combination of the galaxy template (green), an AGN accretion disk component (blue) and an AGN dust torus model (yellow). Top right: PDF for the stellar mass taking into account all possible fractions of AGN emission and providing an upper limit on the stellar mass. So the most probable value (MsPDF, blue dashed line) has higher value than the best-fit stellar mass (MsBEST, red solid line). The 16 and 84 percentile intervals (gray shades) are also indicated. Middle: Emission line fitting of the VIPERS spectrum including the continuum emission (in yellow, top panel), and the broad lines (in blue) decomposed into broad (in red) and narrow (in green) components (zoom-in in the bottom panels). Bottom: Subaru Hyper Suprime-Cam image in the i-band. }
\label{SEDfit}
\end{figure}

In order to select dwarf galaxies among the sample of 1,161 VIPERS BLAGN, we have to derive the stellar masses of the host galaxies. The most common method to derive physical properties is based on the fitting of the spectral energy distribution (SED). 
VIPERS covers the W1 and W4 Canada-France-Hawaii Telescope (CFHT) Legacy Survey Wide (CFHTLS Wide) fields, which provide magnitudes in the optical/infrared bands (u, g, r, i, z). Near-infrared photometry is also available for all the BLAGN, in the K$_{s}$-band from the WIRCam instrument (\citealt{2004SPIE.5492..978P}) at CFHT and in the deeper K$_{video}$ from the VISTA Deep Extragalactic Observations (VIDEO) survey (\citealt{2013MNRAS.428.1281J}). We also include far- and near-ultraviolet measurements from the Galaxy Evolution Explorer (GALEX) satellite (\citealt{2005ApJ...619L...1M}) for 64\% of the BLAGN sample. In addition, 18\% of the BLAGN are observed at mid-infrared wavelengths with Spitzer/IRAC channels centered at 3.6, 4.5, 5.8, and 8.0 $\mu$m and MIPS filters centered at 24, 70, and 160 $\mu$m from the Spitzer WIDE-area Infrared Extragalactic survey (SWIRE) observation of the XMM-Large-Scale Structure (XMM-LSS; \citealt{2004JCAP...09..011P}4) field, which overlaps with the W1 field. Taking advantage of the XMM coverage of the CHTLS W1 field, we use the XMM-XXL catalogue (\citealt{2018A&A...620A..12C}) to identify X-ray detections for 452 out of 1,161 sources in the BLAGN sample. The VIPERS survey was also covered by the NASA's Wide-field Infrared Survey Explorer (WISE; \citealt{2010AJ....140.1868W}) passbands w1, w2, w3, and w4 with effective wavelengths 3.4, 4.6, 12.1, and 22.5 $\mu$m. WISE photometry is available for 45\% of the BLAGN sample. Thanks to the wealth of auxiliary data and multi-wavelength coverage of the VIPERS survey, we were able to perform SED fitting of the BLAGN sample spanning from ultraviolet to infrared wavelengths.
The SED fitting was performed with a modified version of that in \cite{2019ApJ...872..168S}, considering the same SED libraries as in AGNfitter (\citealt{2016ApJ...833...98C}) for the different components of the observed SED, specifically: A grid of stellar population models with exponentially decaying star formation histories with characteristic times ranging from $\tau$=0.1 to 30 Gyr, a \cite{2003PASP..115..763C} initial mass function (IMF), and a model with constant star formation, an AGN accretion disk model, and four AGN dust torus templates depending on the amount of nuclear obscuration in terms of hydrogen column density. The grid of stellar population models ranged from 0.1 to 10 Gyr, with the age of the stellar population capped at the age of the Universe at the redshift of the source. We took into account the dust extinction for both nuclear and galaxy templates with E(B-V) values up to 1 using the reddening law of \cite{1984A&A...132..389P} and \cite{2000ApJ...533..682C}, respectively. A full detailed description of SED model templates is presented in \cite{2019ApJ...872..168S}. 
We first determined the best fit using the $\chi^2$ minimization among all the possible combinations of SED templates. We confirmed that the monochromatic luminosity at rest-frame 2500\AA\ of the best-fitting AGN component correlates with the X-ray luminosity for sources detected in the X-ray, consistent with the observed X-ray-to-ultraviolet correlation for AGN (e.g., \citealt{2016ApJ...819..154L}). 
Since the AGN and galaxy lights are highly degenerated, we then derived a probability distribution function (PDF) for the stellar mass, considering any possible combination of SED parameter space and AGN fraction (f$_\mathrm{AGN}$ = 0 to 1), to evaluate the robust uncertainties taken into account for the degeneracies inherent in the SED fitting. To obtain an upper limit on the best-fit stellar mass (when f$_\mathrm{AGN}$ = 0) we perform the SED fitting with the constraint that the galaxy light dominates over the AGN in the K-band, this is, using the stellar population at the oldest possible age. Since the older stellar populations have higher mass-to-light ratios, this provides a conservative upper limit on the stellar mass. Since when deriving the PDFs all the combinations of AGN-host fractions are taken into account in order to provide an upper limit on the best-fit stellar mass, the most probable value (MsPDF) has higher value than the best-fit stellar mass (MsBEST; see Fig.~\ref{SEDfit}, top right panel). The uncertainties in the stellar mass are then derived as the difference between the highest most probable value (i.e. the upper value of MsPDF) and the best-fit stellar mass (MsBEST). To incorporate the uncertainties caused by differences in the stellar population models, such as the choice of IMF (i.e., M$_\mathrm{*, Salpelter}$ = 1.7 $\times$ M$_\mathrm{*, Chabrier}$), we add an additional 0.2 dex to the uncertainties.
The stellar mass of the \cite{2015ApJ...813...82R} sample was corrected to match the different mass-to-light ratios (\citealt{2020ApJ...889...32S}).  We derived the SFR from the best-fit SED, defined as SFR $\propto e^{t/\tau}$. 

\section{Other stellar-mass measurements}
\label{stellarmass}
Since the stellar masses obtained via SED fitting in galaxies is not unique but can vary from code to code, we also performed the SED fitting using the Code Investigating GALaxy Emission (CIGALE; \citealt{2019A&A...622A.103B}). CIGALE is a state-of-the-art SED fitting tool based on the principles of the energetic balance between dust-absorbed stellar emission in the ultraviolet and optical bands and its re-emission in the infrared. In this work, we used the delayed star formation history module with an optional exponential burst and the \cite{2003MNRAS.344.1000B} single stellar population models assuming the \cite{2003PASP..115..763C} IMF.  The grid of stellar population models ranged from 0.1 to 8 Gyr for the main population and down to 20 Myr for the most recent stellar population. The mass fraction of the recent population varied from 0 up to 0.5. We also utilised the standard nebular emission model from \cite{2011MNRAS.415.2920I} and modeled dust attenuation using flexible models inspired by the \cite{2000ApJ...533..682C} starburst attenuation curve. The reprocessed dust emission was modeled using the dust emission models of \cite{2014ApJ...780..172D}. Lastly, we utilised the AGN emission model by \cite{2006MNRAS.366..767F}. SEDs of these models are then fitted to the VIPERS galaxy SEDs using a Bayesian-like analysis, where the quality of the fit is expressed by the reduced $\chi^2$.  
Using CIGALE we find that the seven BLAGN have stellar masses of log $M_\mathrm{*} \lesssim$ 10 M$_{\odot}$ and thus consistent with low-mass galaxies. The AGN fraction is found to range from 17\% to 42\% with a mean value of 29\%, similarly to the other few AGN dwarf galaxies found at z$>$0.4 (\citealt{2019ApJ...885L...3H}; \citealt{2022MNRAS.516.2736B}).

As AGNfitter, CIGALE can handle many parameters such as star formation history, stellar population models, attenuation law, dust emission and AGN emission. To check how the set of parameters influences the estimated stellar masses, we performed SED fitting adopting different modules. In particular, in each run we changed one of the modules for: i) double exponential star formation history, ii) stellar population models of \cite{2005MNRAS.362..799M}, iii) \cite{1955ApJ...121..161S} IMF, iv) modified \cite{2000ApJ...539..718C} attenuation model, v) \cite{2014ApJ...784...83D} model for modeling dust emission, and vi) Skirtor AGN model (\citealt{2012MNRAS.420.2756S}). The stellar masses vary with different parameters, showing negligible dependence on the choice of dust emission and star formation history up to $\sim$0.2dex overestimation with \cite{1955ApJ...121..161S} IMF and \cite{2005MNRAS.362..799M} stellar populations models. Nevertheless, even with these different parameterizations, the BLAGN still have stellar masses log $M_\mathrm{*} \lesssim$ 10 M$_{\odot}$ consistent with low-mass galaxies. Given that most of the sources have an X-ray detection, we repeated the SED fitting incorporating the X-ray fluxes and using the X-CIGALE code (\citealt{2020MNRAS.491..740Y}). X-CIGALE includes several improvements (X-ray photometry module and a polar dust model) with respect to CIGALE and is commonly used to derive physical properties of AGN and the host galaxy. We used the same modules and parameters as with the CIGALE run except that we used the Skirtor AGN model instead of the one by \cite{2006MNRAS.366..767F}. Using X-CIGALE we find modeled X-ray luminosities in the same range (L$_\mathrm{2-10 keV}$ = 10$^{43}$ - 10$^{44}$ erg s$^{-1}$) as those of the six BLAGN dwarf galaxies with an X-ray detection, which further reinforces the reliability of the SED fitting. 

To obtain an independent measure of the stellar masses, we derived the mass-to-light ratio (M/L) using the Penalized Pixel-Fitting code (pPXF; \citealt{2017MNRAS.466..798C}). pPXF finds the best-fitting spectrum from the grid of stellar spectra templates from the MILES library (\citealt{2010MNRAS.404.1639V}). The grid is generated under assumption of a \cite{1955ApJ...121..161S} IMF using 25 ages uniformly spaced in 7.8 $<$ log [age] (yr) $<$ 10.25 and six metallicities. The templates have a FWHM of 2.51\AA, which we convolved with the wavelength-dependent instrumental dispersion of our data to match the lower spectral resolution of the observed spectra. We are able to fit the seven BLAGN galaxies. We simultaneously fit the gas emission lines (adopting the same kinematics for all of them) and the stellar populations. The fitting includes the emission from the Balmer series, the [O III] and [N II] doublets (with a fixed ratio 1/3), the [O I] doublet (with a fixed ratio 3/1), the [O II] and the [S II]. We compute total stellar masses by multiplication of the stellar M/L calculated in u-,g-,r- and i-bands with the total luminosity in each band. The stellar masses derived from the M/L ratio are in all cases log $M_\mathrm{*} \leq$ 9.7 M$_{\odot}$ and thus consistent with low-mass galaxies. 

\section{Emission line fitting}
\label{broadlines}
To confirm the BLAGN nature of the dwarf galaxies and measure their spectral properties, we fit the optical spectrum using the publicly available Python QSO fitting code (PyQSOFit; \citealt{2018ascl.soft09008G}). The code performs a multi-component spectral fitting by combining host galaxy templates, a dust reddening map, FeII templates, and an input line-fitting parameter list (\citealt{2019ApJS..241...34S}). The code was initially designed to measure spectral properties of SDSS quasars, but it can be tuned to fit VIPERS (and any other) spectra (\citealt{2022MNRAS.516.2736B}). The fitting is performed in the rest-frame and using a $\chi^2$-based method. We first fitted a power-law to the continuum emission by taking a few emission line-free windows around the broad emission lines. After subtracting this continuum emission, we then modeled the H$\alpha$, H$\beta$, H$\delta$, H$\gamma$, and MgII emission lines using a combination of typically one narrow and one broad Gaussian component. A line component is defined as narrow if its FWHM is $<$ 1,200 km s$^{-1}$ (\citealt{2019ApJS..241...34S}). All the lines within a complex were fitted together, but each line complex was fitted separately. For the H$\alpha$ and H$\beta$ complexes, the velocity and width of the narrow components were tied together during the fitting following common procedure. The code measures the continuum luminosity at 3000 and 5100 \AA, and the line flux, FWHM, equivalent width, and dispersion of the broad components. The uncertainty in the line measurements are computed using a MonteCarlo approach. For the seven BLAGN dwarf galaxies SDSS spectra are also available. We fitted these SDSS spectra as detailed above and compared the resulting spectral fitting with the VIPERS one. The SDSS spectra tend to be in general very noisy in the area of emission which we are interested in (mostly due to fringing). We favored the SDSS fit over that of VIPERS only for three sources for which only one emission line is included in the VIPERS spectrum (e.g. only H$\beta$ is in the VIPERS spectrum while in the SDSS there is H$\beta$ and MgII), as the $\chi^2$ of the SDSS fit is in all these cases better than that of the VIPERS one. For two of the three sources for which we use the fit of the SDSS rather than the VIPERS spectrum the continuum luminosity and broad line width are consistent with those of the SDSS quasar catalog of \cite{2019ApJS..244...36C}. 

We note that the 3000 and 5100\ \AA continuum luminosities, which range from 43.8 to 44.6 erg s$^{-1}$, are also consistent with those of the SED fitting at the corresponding frequencies (log $\nu$ (Hz) $\sim$ 15). The spectral fitting for one of the seven sources is shown in Fig.~\ref{SEDfit}, middle panel.

To obtain an independent measurement of the spectral properties, and thus of the black hole masses (see next section), we also fit the VIPERS and SDSS spectra using the code described in Sect. 5.3 of \cite{2015ApJ...815..129S} and which applies the mpfit routine to derive the best-fit parameters and goodness of the fit. As in PyQSOFit, the code fits the continuum with a power-law plus a complex of FeII emission lines in the case of the MgII emission line. The H$\alpha$, H$\beta$, and MgII lines are modeled using a narrow and one or two broad Gaussian components. The fits obtained with only narrow-line components are compared to those with narrow-line and broad Gaussian components, and an F-test is applied to decide whether the additional broad components are needed. The narrow-line components are then subtracted from the spectra to obtain a spectrum with only broad-line components.

\section{Black hole masses}
\label{bhmass}
Black hole masses ($M_\mathrm{BH}$) are estimated assuming that the gas in the broad line region around the black hole is virialized (\citealt{2004ApJ...613..682P}). The velocity of the gas is inferred from the width of the broad H$\beta$ (for z $<$ 0.8) or MgII (for z $>$ 0.8) emission lines, while the line luminosity or rest-frame continuum luminosity at 3000 and 5100 \AA\ is taken as proxy for the size of the broad line region. We use the virial correlations from \cite{2006ApJ...641..689V} and \cite{2012ApJ...753..125S} based on a mean virial factor $\epsilon \sim$1 (e.g., \citealt{2004IAUS..222..109O}; \citealt{2013ApJ...773...90G}). The systematic uncertainties associated with the scatter of these single-epoch virial calibrations are of $\sim$0.3 dex (e.g., \citealt{2008ApJ...673..703M}; \citealt{2012MNRAS.427.3081T}) while the measurement uncertainties are of $\sim$0.1 dex. The total black hole mass uncertainties resulting from summing in quadrature the measurement uncertainties and the systematic uncertainties are of $\sim$0.4 dex. 
From the monochromatic continuum luminosities at 3000 and 5100\ \AA we additionally derive the AGN bolometric luminosities ($L_\mathrm{bol}$) applying the bolometric correction factors of \cite{2006ApJS..166..470R}. This allows us to derive the Eddington ratio as $\lambda_\mathrm{Edd}$ = $L_\mathrm{bol}$/($M_\mathrm{BH} \times 1.3 \times 10^{38}$). 

The black hole masses are derived both from the fits of PyQSOFit and of the routine of \cite{2015ApJ...815..129S}. The results are fully consistent within the uncertainties, with log $M_\mathrm{BH}$ ranging from 7.6 to 8.7 M$_{\odot}$ in the case of PyQSOFit and log $M_\mathrm{BH}$ = 7.7-8.5 M$_{\odot}$ in the case of the Suh et al. routine.

We compute the black hole mass offset ($\Delta M_\mathrm{BH}$) from the $M_\mathrm{BH}$-$M_\mathrm{*}$ correlations using a MonteCarlo approach as in \cite{2018MNRAS.474.1342M}: we assume that the $M_\mathrm{BH}$ and $M_\mathrm{*}$ measurements are independent and that their values and associated uncertainties follow a Gaussian distribution. We assign 100 random variables to the $M_\mathrm{BH}$ and to the $M_\mathrm{*}$ distributions of each source and calculate $\Delta M_\mathrm{BH}$ based on the distribution of 100$^2$ possibilities over the seven BLAGN. The final values of $\Delta M_\mathrm{BH}$ and their error are obtained from the median value and standard deviation, respectively, of the $\Delta M_\mathrm{BH}$ distribution. We also compute the probability (in percentage and $\sigma$) that $\Delta M_\mathrm{BH}$ is larger than zero. We note that the results do not change in a significant manner when increasing the number of random variables to e.g. 500.

\section{Environment}
\label{environment}
The environment of VIPERS galaxies is defined by the local density contrast (see \citealt{2017A&A...602A..15C}) which was measured for four of the seven BLAGN dwarf galaxies. Roughly, the first percentile of the local density distribution for VIPERS galaxies corresponds to void galaxies with an average projected distance to the fifth nearest neighbor D$_{p,5}\sim$3.5$h^{-1}$ Mpc, while the fourth percentile characterizes group and cluster galaxies with an average projected D$_{p,5}\sim$2$h^{-1}$ Mpc. One source is found in the first percentile (low-density environment), one in the second percentile (low-medium density), two in the third percentile (high-medium density), and none in the fourth percentile (high-density environment). This suggests that none of the sources is located in a cluster or group of galaxies, although they may preferably reside in environments denser than voids (though not reaching the dimensions of clusters or groups). A larger sample would be needed to confirm such a tendency. We note that a statistical study of the environment of AGN dwarf galaxies out to z$\sim$0.9 was recently performed by \cite{2023MNRAS.518..724S}, finding that dwarf galaxies prefer low-density environments independently of whether they host an AGN or not. This is in agreement with the absence of a significant dependence between AGN activity and environment found for dwarf galaxies at lower redshifts (e.g., \citealt{2018ApJ...861...50B}; \citealt{2020MNRAS.498.4562M}; \citealt{2020MNRAS.496.2577K}; \citealt{2022MNRAS.511.4109D}). 
Moreover, none of the seven VIPERS sources is located in the SDSS galaxy group catalog of \cite{2021ApJ...923..154T}.

We also thoroughly study the Subaru Hyper Suprime-Cam i-band images of each of the seven BLAGN dwarf galaxies. For this we download cutouts of semi-width and semi-height of 0.02 deg centered on the VIPERS sources and search for galaxies at a similar redshift to ours by making using of extragalactic catalogs such as SDSS. We also look for merger signatures, such as interactions or tidal tails. In none of the VIPERS galaxies is there evidence for the presence of a companion or pair. All the sources observed close to our galaxies are either background sources at a different redshift, are not catalogued, or are foreground stars.




\end{document}